\documentclass[a4paper,12pt]{article}
\usepackage[cp1251]{inputenc}
\usepackage[dvips]{graphicx}
\graphicspath{{.}}

\begin{document}
%\Large
 \mathsurround=2pt \sloppy
%\begin{center}
\title{\bf Anomalous temperature dependence of the order parameter of
            a superconductor with weakly correlated impurities}
\author{I. A. Fomin \\
{\it P. L. Kapitza Institute for Physical Problems},\\{\it
Kosygina 2,
 119334 Moscow, Russia};\\ {\it Moscow Institute of Physics and
Technology}, \\{\it Dolgoprudny, Mosow region}}
%\end{center}
\maketitle
\begin{abstract}
It is shown that weak correlations between pair-breaking
impurities in superconductors influence the temperature dependence
of the order parameter within the Ginzburg and Landau region if
correlation radius of impurities $R$ is greater than the coherence
length of the superconductor $\xi_0$. Dependence of a square of
the average order parameter on the temperature difference $T_c-T$
changes its slope in a region $\xi_0\sqrt{T_c/(T_c-T)}\sim R$.
Influence of correlations of impurities on other thermodynamic
properties of superconductors is discussed.
\end{abstract}

   Impurities make the condensate of  Cooper pairs spatially
nonuniform. Manifestation of this non-uniformity is particularly
strong in unconventional superconductors and in conventional
superconductors with magnetic impurities.  In a vicinity of the
transition temperature $T_c$ free energy of such superconductor
can be written as  Ginzburg and Landau functional with
coefficients, which are random functions of coordinate. For a
scalar order parameter $\Psi(\textbf{r})$
$$
\emph{F}_s\{\Psi(\textbf{r})\}=\emph{F}_n+
\int\{a(\textbf{r})|\Psi(\textbf{r})|^2+
\frac{1}{2}b(\textbf{r})|\Psi(\textbf{r})|^4+
c(\textbf{r})|\nabla\Psi(\textbf{r})|^2\}d^3r.           \eqno(1)
$$

According to the analysis of Larkin and Ovchinnikov \cite{LO} the
most strong effect on the average order parameter have
fluctuations of the coefficient $a(\textbf{r})$ i.e. fluctuations
of the local transition temperature $T_c(\textbf{r})$:
$a(\textbf{r})=\alpha(T-T_c(\textbf{r}))$. Because of these
fluctuations the temperature dependence of the average order
parameter $\langle\Psi\rangle$ deviates from the linear dependence
$\langle\Psi\rangle^2\sim(T_c-T)/T_c$ characteristic of the
uniform superconductor and becomes singular in a narrow interval
below the superfluid transition temperature
$(T_c-T)/T_c\sim(\xi_0/l)^4(1/(n\xi_0)^2)$, where $l$ is the mean
free path and $n$ - the density of impurities. The linear
dependence is preserved outside of this region. This result is
obtained under assumption that impurities are not correlated. It
has been shown recently, that correlations with a radius $R$ which
is greater than $\xi_0$  strongly affect suppression of $T_c$ by
impurities \cite{f1,f2}. Such situation is realized for the
superfluid $^3$He in aerogel. Experimental data for thermodynamic
properties of superfluid $^3$He in aerogel, such as  the square of
the Leggett frequency $\Omega_L^2$ and $\rho_s$ \cite{Dm,PD}
deviate from the linear dependence on $T_c-T$ in a much wider
interval of temperatures than that, estimated on a basis of
Ref.\cite{LO}. The observed dependence of these quantities
 bends upward when $T_c-T$ increases. The aim of this paper is to
show that this anomaly can be interpreted as the effect of
correlations among the impurities. The effect of correlations is
general in a sense, that presence and a character of deviations do
not depend on a particular form of the order parameter. For
demonstration of the effect the simplest example of
 a superconductor with the scalar order parameter
$\Psi(\textbf{r})$ will be considered.

Let us follow the argument of Ref.\cite{LO} with the modifications
required by the presence of correlations. Free energy (1) can be
rewritten in terms of
 the dimensionless order parameter
$\psi(\textbf{r})=\Psi(\textbf{r})/\Psi_0$, where $\Psi_0$ is the
absolute value of the order parameter for pure superconductor at
$T=0$ obtained by extrapolation of linear dependence of $|\Psi|^2$
on $(T_{c0}-T)$ from the transition temperature of the pure
superconductor $T_{c0}$:
$$
\emph{F}_s\{\psi(\textbf{r})\}=\emph{F}_n+ T_{c0}\Delta
c_0\int\{[\tau-\eta(\textbf{r})]|\psi(\textbf{r})|^2+
\frac{1}{2}|\psi(\textbf{r})|^4+
\xi_s^2|\nabla\psi(\textbf{r})|^2\}d^3r. \eqno(2)
$$
Other notations here are: $\Delta c_0$ -- the specific heat jump
in the pure superconductor, $\tau=(T-T_{c1})/T_{c0}$,
$T_{c1}=\langle T_c(\textbf{r})\rangle$, so that the relative
fluctuation of local transition temperature
$\eta(\textbf{r})=(T_c(\textbf{r})-T_{c1})/T_{c0}$,  it vanishes
after averaging. The elasticity coefficient
$\xi_s^2=\frac{7\zeta(3)}{12}\xi_0^2$ in the BCS theory. In these
notations Ginzburg and Landau equation reads as:
$$
[\tau-\eta(\textbf{r})]\psi+\psi|\psi|^2-\xi_s^2\Delta\psi=0
   \eqno(3)
$$
At $T<T_c$ the long-range order is established, characterized by
the average order parameter $\langle\psi\rangle $. The angular
brackets denote averaging over ensemble of impurities.
 Solution of Eq. (3) can be sought in a form
$\psi(\bf{r})=\langle\psi\rangle(1+\chi({\bf r}))$. When
$\eta({\bf r})$ is small $\chi({\bf r})$ is small too, except for
the mentioned above temperature region close to $T_c$, where
$\langle\chi({\bf r})\chi({\bf r})\rangle$ diverges. Averaging
Eq.(3) over ensemble of impurities renders expression for
$\langle\psi\rangle^2$ in terms of the average products
$\langle\eta\chi\rangle$ and $\langle\chi\chi\rangle$:
$$
\langle\psi\rangle^2=\frac{\langle\eta\chi\rangle-\tau}{1+3\langle\chi\chi\rangle}
      \eqno(4)
$$
The fluctuating part $\chi(\bf{r})$ can be found from the
linearized equation (2):
$$
(\tau-\eta({\bf r})+3\langle\psi\rangle^2)\chi-\xi_s^2\Delta\chi=
\eta({\bf r})-\langle\eta\chi\rangle.           \eqno(5)
$$
Its solution can be formally written in terms of the Green
function $G({\bf r},{\bf r'})$:
$$
\chi({\bf r})=\int G({\bf r},{\bf r'})(\eta({\bf r'})-
\langle\eta\chi\rangle)d^3r'.                       \eqno(6)
$$
Re-writing this solution in the momentum representation we can
find the averages, entering Eq. (4):
$$
\langle\eta(\textbf{r})\chi(\textbf{r})\rangle=\int\langle\eta(\textbf{-k})
G({\bf k},{\bf k'})\eta({\bf k'})\rangle\frac{d^3k}{(2\pi)^3}
\frac{d^3k'}{(2\pi)^3}\left[1+ \int\langle\eta({\bf -k}) G(\textbf
{k},0)\rangle\frac{d^3k}{(2\pi)^3}\right]^{-1} \eqno(7)
$$
The average in the numerator can be found by term-by-term
averaging of the diagrammatic series Fig.1.

\begin{figure}
\begin{center}
\includegraphics[width=0.75\linewidth,keepaspectratio]{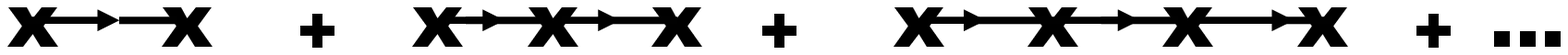}
\end{center}
\caption ~
\end{figure}

To every arrow here corresponds the unperturbed Green function
$G_0({\bf k},{\bf k'})=(2\pi)^3\delta({\bf k}-{\bf
k'})[\tau+3\langle\psi\rangle^2+\xi_s^2k^2]^{-1}$ and to a cross
$\eta(\textbf{k}_1-\textbf{k}_2)$, $\textbf{k}_1$ is the in-coming
and $\textbf{k}_2$ -- the out-going momenta. Comparison of the
averaged diagrammatic series Fig.1 with that for the average Green
function relates $\langle\eta G\eta\rangle$ to $\langle G\rangle$:
$$
\langle\eta G\eta\rangle=G_0^{-1}[\langle G\rangle G_0^{-1}-1],
\eqno(8)
$$
which in its turn can be expressed in terms of the self-energy
$\Sigma(\textbf{k},\tau)$:
$$
\langle
G(\textbf{k})\rangle=[\tau+3\langle\psi\rangle^2-\Sigma(\textbf{k},\tau)+\xi_s^2k^2]^{-1}
  \eqno(9)
$$
Using analogous argument for evaluation of $\langle\eta G\rangle$
we find eventually that
$$
\langle\chi\eta\rangle=\Sigma(0,\tau).             \eqno(10)
$$
In a principal order on the perturbation $\eta(\textbf{r})$
$$
\Sigma(0,\tau)=\int\frac{\langle\eta(\textbf{-k}_1)\eta(\textbf{k}_1)\rangle}
{\tau+3\langle\psi\rangle^2-\Sigma(\textbf{k}_1,\tau)+\xi_s^2k_1^2}
\frac{d^3k_1}{(2\pi)^3}.    \eqno(11)
$$
Of practical interest is a situation when
$\eta(\textbf{r})=\sum_a\eta^{(1)}(\textbf{r}-\textbf{r}_a)$,
where $\eta^{(1)}(\textbf{r}-\textbf{r}_a)$ is a contribution of
individual impurity  situated at the position $\textbf{r}_a$. In
that case the correlation function
$\langle\eta(\textbf{-k})\eta(\textbf{k})\rangle=
n|\eta^{(1)}(\textbf{k})|^2S(\textbf{k})$, where
$S(\textbf{k})=\langle\sum_b\exp[i\textbf{k}(\textbf{r}_b-\textbf{r}_a)]\rangle$
is the structure factor. For non-correlated impurities only one
term with $\textbf{r}_b=\textbf{r}_a$ contributes to the sum.
Effect of correlations is contained in
$\bar{S}(\textbf{k})=S(\textbf{k})-1$, which is Fourier transform
of the correlation function in the coordinate space
$v(\textbf{r})$. When impurities are correlated on a distance
$\sim R$ the $\bar{S}(\textbf{k})$ is enhanced for $k\sim 1/R$.
For estimation of the effect of correlations
 a model expression  ("$\delta$-model"\cite{f2}):
$\bar{S}(\textbf{k})=2\pi^2R^2nv(0)\delta(k-1/R)$ can be used. In
particular, one can show that corrections to the principal
expression for $\Sigma(0,\tau)$, given by Eq.(11) contain extra
factor $R^2/\xi_0l\equiv \varepsilon$. In what follows we assume,
that $\varepsilon\ll 1$ and treat effect of correlation as a
perturbation. In a principal order on $\varepsilon$ in the
denominator of the integrand in Eq. (11) we can set
$\langle\psi\rangle^2=\Sigma(0,\tau)-\tau$ and take
$\Sigma(\textbf{k},\tau)\approx\Sigma(0,\tau)$. Then Eq. (11)
determines a function $\Sigma(0,\tau)$. In this equation a
combination $u=\Sigma(0,\tau)-\tau$ is taken as a new variable. In
the expression (11) for $\Sigma(0,\tau)$  we separate part which
is finite at $u=0$.
 As was discussed before \cite{f1,f2}, it determines the second
order correction to the shift of $T_c$. The new transition
temperature $T_{c2}$ is then
$T_{c2}=T_{c1}+T_{c0}\int\frac{n|\eta^{(1)}(\textbf{k}_1)|^2S(\textbf{k}_1)}
{\xi_s^2k_1^2}\frac{d^3k_1}{(2\pi)^3}$. It is convenient to
introduce a variable $t$, which counts temperature from $T_{c2}$:
$t=(T-T_{c2})/T_{c0}$. In these notations Eq.(11) renders relation
between $u$ and $t$:
$$
u\left[1+2\int\frac{n|\eta^{(1)}(\textbf{k}_1)|^2S(\textbf{k}_1)}
{\xi_s^2k_1^2[2u+\xi_s^2k_1^2]}\frac{d^3k_1}{(2\pi)^3}\right]=-t.
\eqno(12)
$$
In terms of $u$
$\langle\psi\rangle^2=u/(1+3\langle\chi\chi\rangle)$. When
$S(\textbf{k}_1)$ is strongly peaked at $k_1\sim 1/R$ the
dependence of $u$ on $t$ changes its slope in a region
$u\sim(\xi_s/R)^2$, i.e. when $\xi(T)\sim R$ and $\xi(T)$ is
defined as $\xi_0/\sqrt{u}$.  At $u\gg(\xi_s/R)^2$ asymptotically
$u\approx(T_{c1}-T)/T_{c0}$, i.e. it depends linearly on a
distance from the average transition temperature $T_{c1}$. In the
opposite limit $u\ll(\xi_s/R)^2$ u is linear in
$t=(T_{c2}-T)/T_{c0}$, with a different slope. The relative change
of the slope is on the order of $\varepsilon^2$. Using for
$\bar{S}(\textbf{k})$ the $\delta$-model we have:
$$
u=\frac{t}{1+2n^2|\eta^{(1)}(0)|^2v(0)(R/\xi_s)^4}. \eqno(13)
$$
If impurities, like aerogel, have a tendency to form clusters,
$v(0)>0$ and the slope of $u(t)$ at $u\ll(\xi_s/R)^2$ is smaller
than at $u\gg(\xi_s/R)^2$, so the $u(t)$ bends upward. Together
with $u(t)$ changes its slope
$\langle\psi\rangle^2(t)=u/(1+3\langle\chi\chi\rangle)$. The
average $\langle\chi\chi\rangle$ in the denominator is on the
order of $\varepsilon^2$:
$$
\langle\chi\chi\rangle=\int\frac{n|\eta^{(1)}(\textbf{k})|^2S(\textbf{k})}
{[2u+\xi_s^2k^2]^2}\frac{d^3k}{(2\pi)^3}.       \eqno(14)
$$
This correction does not influence asymptotic of the dependence of
$\langle\psi\rangle^2$ on $t$ at $u\gg(\xi_s/R)^2$, but at
$u\ll(\xi_s/R)^2$ it increases the change of the slope. Due to
this correction physical quantities, which depend on averages of
different powers of the order parameter will have different
changes of the slope. E.g. the NMR frequencies are proportional to
$\langle\psi^2\rangle=\langle\psi\rangle^2(1+\langle\chi\chi\rangle)\simeq
u/(1+2\langle\chi\chi\rangle)$. For the model
$\bar{S}\sim\delta(k-1/R)$ at $u\ll(\xi_s/R)^2$:
$$
\langle\psi^2\rangle=\frac{t}{1+4n^2|\eta^{(1)}(0)|^2v(0)(R/\xi_s)^4}.
    \eqno(15)
$$

Further thermodynamic properties can be found as derivatives of
the average free energy over $T$. Using the expression
$\emph{F}_s-\emph{F}_n=-\frac{T_{c0}\Delta
c_0}{2}\langle\int|\psi|^4d^3r\rangle$ and relation
$\langle|\psi|^4\rangle=\langle\psi\rangle^4(1+6\langle\chi\chi\rangle)$
we arrive at:
$$
\emph{F}_s-\emph{F}_n=-\frac{T_{c0}\Delta C_0}{2} u^2, \eqno(16)
$$
i.e. a gain of the free energy is proportional to $u^2$ and not to
$t^2$. The specific heat jump, following from Eq. (16) is:
$$
\Delta C_i=\Delta C_0
T_{c0}T_{c2}\left(\frac{du}{dt}\right)^2_{T\to T_{c2}}. \eqno(17)
$$
For the $\delta$-model $\Delta C_i=\Delta C_0
\frac{T_{c2}}{T_{c0}}(1+2n^2|\eta^{(1)}(0)|^2v(0)(R/\xi_s)^4)^{-2}$.
There is extra suppression of the jump due to correlations.

  So, we conclude, that the anomaly of temperature dependence of
thermodynamic properties of superfluid $^3$He in aerogel is
qualitatively the same as that found in the considered example of
a superconductor with a one-component order parameter and
correlated impurities. The quantitative comparison of the obtained
results with the data for $^3$He would not have sense, because
superfluid $^3$He has a multi component order parameter. This
opens a possibility for correlated impurities to interact with
several different modes of fluctuations of the order parameter.
For a quantitative description of the anomaly all of these modes
have to be taken into account. Nevertheless a qualitative
estimation of the correlation radius based on the relation
$\xi(T)\approx R$ at a temperature of a change of the slope is
close to other estimations.

Because of its universal character the anomaly can occur in
superconductors with different order parameters and can be
considered as an indication that impurities are correlated and
correlation radius is greater than $\xi_0$. A distance from $T_c$
at which dependence of $\langle\psi\rangle^2$ on $T_c-T$ changes
its slope renders estimation of the correlation radius.

I thank E.V. Surovtsev for useful discussions. This work was
supported in part by RFBR grant \# 09-02-12131 ofi-m.

\end{document}